\documentclass[aps,pre,showpacs,groupedaddress]{revtex4}
\usepackage{amsfonts,amssymb,amsmath,amsbsy}
\usepackage{epsfig}

\begin{document}

\title{Code optimization, frozen glassy phase and improved decoding algorithms for low-density parity-check codes}

\author{Haiping Huang}
 \affiliation{Department of Computational
Intelligence and Systems Science, Tokyo Institute of Technology,
Yokohama 226-8502, Japan}

\date{\today}

\begin{abstract}
The statistical physics properties of low-density parity-check codes
for the binary symmetric channel are investigated as a spin glass
problem with multi-spin interactions and quenched random fields by
the cavity method. By evaluating the entropy function at the
Nishimori temperature, we find that irregular constructions with
heterogeneous degree distribution of check (bit) nodes have higher
decoding thresholds compared to regular counterparts with
homogeneous degree distribution. We also show that the instability
of the mean-field calculation takes place only after the entropy
crisis, suggesting the presence of a frozen glassy phase at low
temperatures. When no prior knowledge of channel noise is assumed
(searching for the ground state), we find that a reinforced strategy
on normal belief propagation will boost the decoding threshold to a
higher value than the normal belief propagation. This value is close
to the dynamical transition where all local search heuristics fail
to identify the true message (codeword or the ferromagnetic state).
After the dynamical transition, the number of metastable states with
larger energy density (than the ferromagnetic state) becomes
exponentially numerous. When the noise level of the transmission
channel approaches the static transition point, there starts to
exist exponentially numerous codewords sharing the identical
ferromagnetic energy.
\end{abstract}

 \maketitle

\section{Introduction}
In modern wireless communication, reliable transmission of
information in a noisy environment can be achieved, proved by
Shannon who put forward the celebrated channel encoding
theorem~\cite{Shannon-1948a,Shannon-1948b}. This theorem states that
the error-free transmission is possible as long as the code rate $R$
(the ratio between the number of bits in the original message and
the number of bits in the transmitted message) doesnot exceed the
capacity of the channel (Shannon's bound).
 The relation between spin glass models and information theory was
first well established by Sourlas in 1989~\cite{Sourlas-1989}. After
that, statistical physics methods especially replica trick was
applied to analyze the typical properties of coding and decoding
problems~\cite{Nishimori-Wong-1999,Kanter-1999,Kaba-2000a,Montanari-2001,Tanaka-2003jpa,Migliorini-2006,Mora-pre06,Neir-2008,Huang-2009},
and the dynamical properties of decoding
process~\cite{Franz-2002pre}. Studies of this line over decades have
achieved significant results, some of which are not able to be easily
obtained using traditional methods of information theory, and all
these results are remarkably consistent with those of information
theory~\cite{RR-2008}.

The channel encoding theorem does not tell us how to construct an
optimal code that saturates the Shannon's bound. Information theory
community have devoted lots of efforts to devise (near) optimal
codes over last several decades~\cite{Mackay-1999}. Codes of
Gallager type are promising candidates since they have vanishing bit
error rate and can saturate the Shannon limit. Gallager type
error correcting code (also known as low-density parity-check (LDPC) code) was first discovered in
1962~\cite{Gallager-1962}, but was abandoned soon due to the limited
computational ability at that time. This code was rediscovered by
Mackay and Neal in 1996~\cite{Mackay-1996}. Since then, the LDPC codes were extensively studied in
either construction schemes or decoding algorithms. Methods of
statistical physics, complementary to those used in information
theory, enable one to attain a complete picture of decoding
process by analyzing global properties of the corresponding free
energy landscape. They also allow one to optimize the performances
of various codes by changing some construction parameters. Additionally, the
coding and decoding process can be mapped onto the factor
graph~\cite{MM-2009} (also called Tanner graph~\cite{RR-2008}) with
locally tree-like structure, which facilitates the statistical
mechanics analysis.

The known picture for Gallager-type codes
is~\cite{Montanari-2001,Franz-2002pre}: for sufficiently small noise
levels in a transmission channel, the ferromagnetic solution is the
only stable solution and the complete decoding is possible. Simple
local search algorithm can recover the corrupted bits. However, as
the noise level increases up to the spinodal or dynamical transition
point where an exponentially large number of metastable states
appear (these states are suboptimal ferromagnetic solutions and hide
the original message), decoding algorithms finally fail to identify
the solution in available time scales. Recent studies using one step
replica symmetry breaking theory showed that the theoretical
decoding limit can be pushed to a higher
value~\cite{Migliorini-2006}. This implies that in this
computationally hard region~\cite{NPC-1978}, detailed study of
codeword space structure and even metastable state landscape is
needed, which might suggest novel efficient decoding schemes and
insights towards the glassy dynamics of local search heuristics.
When the noise level crosses the thermodynamic (static) transition
point, a fraction of the metastable states are degenerate with the
true codeword, and error-free communication becomes impossible. This
transition point is upper-bounded by the Shannon limit.

In this paper, we provide several physical insights for
understanding the LDPC codes. First, by tuning the construction
parameter, we compute the \textit{entropy} value at the Nishimori
temperature~\cite{Nishimori-1993}(equivalently we have the prior
knowledge of the channel statistics, e.g., the noise level), which
reveals that the irregular constructions where degree of nodes in
the factor graph follows a distribution are able to tolerate higher
noise levels for reliable transmission, compared with the regular
counterparts with fixed degree of nodes in the factor graph.

 Second,
to probe the geometrical structure of codewords, we also compute the
free energy as a function of temperatures. It is found that the
\textit{entropy crisis} (the free energy takes a maximal value at a
finite temperature) occurs before the instability of the mean-field
calculation, which shows the existence of a frozen glassy phase at
low temperatures~\cite{Montanari-2001,Martin-2005}. At zero
temperature, the decoding process amounts to searching for the
ground state. In this situation, in the presence of high enough
noise level, long-range correlation develops and the first-level
assumption (replica symmetric approximation) breaks down,
consequently, one has to adopt replica symmetry breaking scenario as
a better approximation, under which the complexity of metastable
states (growth rate of the number of metastable states with the
system size) is computed, implying that the number of metastable
states with higher energy starts to grow exponentially when the
dynamical transition is approached. When the noise level of the
channel exceeds the critical point, even the number of states
sharing the same energy with the unique true ferromagnetic state,
starts to proliferate exponentially.

Finally, we observe divergence of local fields in the glassy regime,
consistent with the frozen picture of codeword space structure.
Furthermore, we show that a reinforced belief propagation (rBP) can
\textit{improve} the decoding performance at zero temperature,
although the same scheme has little effects on optimal decoding (at
Nishimori temperature). The associated decoding threshold almost
coincides with the dynamical transition predicted by the theory. The
frozen codeword picture explains the algorithmic hardness for the
improved algorithm~\cite{Zecchina-2008entropy}.

The rest of this paper is organized as follows. The low-density
parity-check code and the associated spin glass model are introduced
in Sec.~\ref{sec_model}. In Sec.~\ref{sec_MFC}, we compute typical
value of the free energy function by using the cavity
method~\cite{MM-2009}, and derive the mean-field formulae for the
entropy function at the Nishimori temperature, under the replica
symmetric (RS) ansatz. Furthermore, we derive the one-step replica
symmetry broken ($1$RSB) solution for the current problem when RS
ansatz becomes incorrect at zero temperature. In this section, we
also present a reinforced belief propagation algorithm for improving
zero temperature decoding performance on single instances. In
Sec.~\ref{sec_result}, the numerical simulation results on single
instances and the theoretical prediction of RS and $1$RSB
approximations are obtained and discussed. We give final remarks and
summary in Sec.~\ref{sec_Con}. Details of the numerical method to
solve $1$RSB equations in Sec.~\ref{subsec:ZTD} are collected in
Appendix~\ref{sec:appendix}.

\section{Model}
\label{sec_model} The information transmission process in modern
wireless communication can be formulated as a channel coding
problem~\cite{Shannon-1948a, RR-2008}, in which a message of length
$N$ is transformed into a redundant transmitted message of length
$M(>N)$. The encoded message is called codeword. We assume each
entry of the message takes Boolean value $0,1$. The original message
is denoted by $\boldsymbol{\xi}$, while the transmitted message by
$\mathbf{t}$. The encoding process is completed by taking
$\mathbf{t}=\mathbf{G}^{T}\boldsymbol{\xi}$~\cite{Gallager-1962,Mackay-1999},
where
$\mathbf{G}=\left[\mathbf{I}|(\mathbf{C}_{2}^{-1}\mathbf{C}_{1})^{T}\right]$,
i.e., a concatenation of two matrixes. $\mathbf{I}$ is an $N\times
N$ identity matrix. $\mathbf{G}$ is chosen such that
$\mathbf{H}\mathbf{G}^{T}=0$ with the parity-check matrix
$\mathbf{H}=\left[\mathbf{C}_{1}|\mathbf{C}_{2}\right]$.
$\mathbf{C}_{1}$ and $\mathbf{C}_{2}$ are $(M-N)\times N$ and
$(M-N)\times(M-N)$ sparse matrixes respectively. Upon encoding, the
message $\mathbf{t}$ is transmitted through a noisy channel which we
assume binary symmetric and memoryless, i.e., the channel is
characterized by the following probability:
\begin{equation}\label{BSC}
    p_{n}(\zeta_{i})=(1-p)\delta(\zeta_{i})+p\delta(\zeta_{i}-1),
\end{equation}
where the noise level $p$ is the flip rate of the channel. The received message is
corrupted by the noise as
$\mathbf{r}=\mathbf{t}+\boldsymbol{\zeta}$, in which each bit is
flipped independently by the noise. The symmetry property of the
channel means that the conditional probability
$P(r_{i}=0|t_{i}=0)=P(r_{i}=1|t_{i}=1)=1-p$, and
$P(r_{i}=0|t_{i}=1)=P(r_{i}=1|t_{i}=0)=p$.  For the binary symmetric
channel (BSC), the Shannon bound is expressed as $R_{c}=1-H_{2}(p)$
where $H_{2}(p)=-p\log_{2}p-(1-p)\log_{2}(1-p)$ is the binary
entropy. The decoding is carried out by calculating the so-called
syndrome vector $\mathbf{z}$~\cite{RR-2008}:
\begin{equation}\label{synd}
\mathbf{z}=\mathbf{H}\mathbf{r}=\mathbf{H}\boldsymbol{\zeta}.
\end{equation}
An estimate of the original message $\boldsymbol{\xi}$ is then
obtained as the first $N$ bits of $\mathbf{r}+\mathbf{n}$, where
$\mathbf{n}$ is the estimate of the true noise vector
$\boldsymbol{\zeta}$, obtained by solving the parity check
equations~\cite{Kaba-2002}:
 \begin{equation}\label{PCE}
\mathbf{z}=\mathbf{H}\mathbf{n}.
\end{equation}
All the above matrix computation is based on ${\rm mod}$ $2$
addition. To define our mean-field model, we transformed the Boolean
variable $n_{i}$ into the Ising one $\sigma_{i}$ via
$\sigma_{i}=(-1)^{n_{i}}$, and thus the ${\rm mod}$ $2$ addition
corresponds to a product. In the remaining part, the noise
$\zeta_{i}$ becomes an Ising variable as well.

\begin{figure}
\centering
    \includegraphics[bb=154 580 398 751,width=0.5\textwidth]{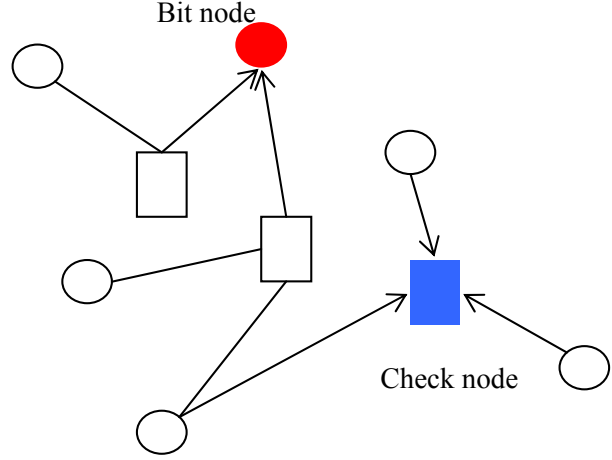}
  \caption{
  (Color online) Factor graph representation of LDPC codes. The
  graph consists of two kinds of nodes (bit node and check node).
  The arrow shows the passing messages (cavity probabilities) to
  compute the free energy contribution of adding one bit node or one
  check node (the full nodes).
  }\label{FactGraph}
\end{figure}

Introducing an inverse temperature $\beta$ as a control parameter,
the posterior probability of a spin configuration
$\boldsymbol{\sigma}$ is given by the Bayes theorem~\cite{Kaba-2004}:
\begin{equation}\label{poster}
  P(\boldsymbol{\sigma}|\mathbf{z})=\frac{\exp\bigl(-\beta \mathcal{H} (\boldsymbol{\sigma})\bigr)}
    {Z},
\end{equation}
where $Z$ is the partition function depending on the channel
statistics and code constructions. The Hamiltonian is then defined
as
\begin{equation}\label{Ham}
    \mathcal
    {H}(\boldsymbol{\sigma})=-\gamma\sum_{\mu=1}^{M-N}\left(\prod_{i\in\partial\mu}\sigma_{i}-1\right)-F\sum_{i=1}^{M}\zeta_{i}\sigma_{i},
\end{equation}
where $\partial\mu$ denotes the neighbors of check $\mu$, and a
gauge transformation $\sigma_{i}\rightarrow\sigma_{i}\zeta_{i}$ has
been made~\cite{Murayama-2000}, which leads to the presence of
random fields in the last term. The magnitude of the random field is
obtained from the priori probability of the noise as
$F=\frac{1}{2}\ln\frac{1-p}{p}$. $\gamma$ will be finally sent to
infinity to enforce $M-N$ parity-check constraints in
Eq.~(\ref{PCE}). $\partial\mu$ denotes the set of non-zero entries
in $\mu$-th row of the parity check matrix $\mathbf{H}$. Note that
the above code construction implies that $\mathbf{H}$ is an
$(M-N)\times M$ sparse matrix, whose total number of non-zero
entries in each row $k$ and that in each column $l$ follow the
degree profile $(\mathcal {P},\Lambda)$ where
$\Lambda(x)=\sum_{l}\Lambda_{l}x^{l}$ and $\mathcal
{P}(x)=\sum_{k}\mathcal {P}_{k}x^{k}$. In this context, we define
the mean field model on a random factor graph in which the mean node
(bit) degree $\left<l\right>=\Lambda'(1)$ and the mean function node
(check) degree $\left<k\right>=\mathcal {P}'(1)$~\cite{Richa-2001b}.
As shown in fig.~\ref{FactGraph}, the factor graph is a bipartite
graph on which there are two kinds of nodes: one is the variable
node (bit) and the other is the function node (check). An edge joins
a variable node $i$ and a function node $\mu$ if and only if the bit
$i$ is involved in $\mu$-th parity check equation. We can also
deduce the edge-perspective degree~\cite{MM-2009} profile
$(\lambda_{l}=l\Lambda_{l}/\left<l\right>,\rho_{k}=k\mathcal
{P}_{k}/\left<k\right>)$, specifying the probabilities that a
randomly selected edge is connected to a node of degree $l$ and to a
function node of degree $k$, respectively. The regular code is
defined when only one $\Lambda_{l}$ ($\Lambda_{k}$) is non-zero,
then the code rate can be obtained by $R=N/M=1-l/k$, where $l$ and
$k$ are replaced by their mean value for irregular
codes~\cite{Luby-2001}.

\section{Mean-field Computation}
\label{sec_MFC}

To study the free-energy landscape of error-correcting codes, we
define the following partition function:
\begin{equation}\label{CPT}
    Z=\sum_{\boldsymbol{\sigma}}e^{-\beta\mathcal {H}(\boldsymbol{\sigma})}
\end{equation}
where $\mathcal {H}(\boldsymbol{\sigma})$ is defined in
Eq.~(\ref{Ham}). In the thermodynamic limit, the entropy density $s$
can be computed via:
\begin{equation}\label{entro}
s=\frac{1}{M}\ln Z-\frac{\beta F}{M}\sum_{i}\zeta_{i}m_{i}.
\end{equation}
 The free energy density
$f\equiv-T\ln Z/M$ and the magnetization
$m_{i}\equiv\left<\sigma_{i}\right>_{P(\boldsymbol{\sigma}|\mathbf{z})}$
will be calculated under the mean field approximation in the
following sections.

\subsection{Replica symmetric approximation}
\label{subsec:RS} We first derive the formula under the replica
symmetric approximation~\cite{MM-2009,Huang-2009}. The key idea is
that, in a modified graph where a bit node $i$ is removed, then all
the incoming probabilities $\hat{p}_{\mu\rightarrow i}(\sigma_{i})$
become independent with each other, which is reasonable when the
graph is sparse and the glass transition does not happen (there are
no long-range correlations on the graph). $\hat{p}_{\mu\rightarrow
i}(\sigma_{i})$ denotes the probability that a check constraint
$\mu$ is satisfied given the value of $\sigma_i$. Thus one can write
down the free energy contribution of a bit node (see
fig.~\ref{FactGraph}):
\begin{equation}\label{dFi}
    e^{-\beta\Delta f_{i}}=\sum_{\sigma_{i}}e^{\beta
    h_{i}\sigma_{i}}\prod_{\mu\in\partial i}\hat{p}_{\mu\rightarrow
    i}(\sigma_{i}),
\end{equation}
where $\partial i$ denotes the neighbors of bit $i$, and the
external field $h_{i}=F\zeta_{i}$. Likewise, in a modified graph
where a function node $\mu$ is removed, then all the incoming
probabilities $p_{i\rightarrow \mu}(\sigma_{i})$ are also
independent with each other.  $p_{i\rightarrow \mu}(\sigma_{i})$
denotes the probability that bit $i$ takes value $\sigma_i$ in the
absence of check $\mu$. Thus the free energy contribution of a
function node (see fig.~\ref{FactGraph}) can be expressed as:
\begin{equation}\label{dFa}
    e^{-\beta\Delta f_{\mu}}=\sum_{\{\sigma_{i}\}:i\in\partial\mu}\delta_{\prod_{i\in\partial\mu}\sigma_{i},1}\prod_{i\in\partial\mu}p_{i\rightarrow
    \mu}(\sigma_{i}),
\end{equation}
where $\delta$-function ensures that the parity check constraint
$\mu$ is satisfied. Interested readers can find details about
Eqs.~(\ref{dFi}) and~(\ref{dFa}) in Ref.~\cite{Huang-2009} for the
cavity method applied to error correcting codes. Finally, one can
parameterize these two probabilities as
$p_{i\rightarrow\mu}(\sigma_{i})=\frac{1+m_{i\rightarrow\mu}\sigma_{i}}{2}$
and $\hat{p}_{\mu\rightarrow
i}(\sigma_{i})\frac{1+\hat{m}_{\mu\rightarrow i}\sigma_{i}}{2}$, and
get the free energy density
\begin{equation}\label{freeEnergy}
    f=\frac{1}{M}\sum_{i}\Delta f_{i}-\frac{1}{M}\sum_{\mu}(k_{\mu}-1)\Delta
    f_{\mu},
\end{equation}
where $k_{\mu}$ is the degree of function node $\mu$ and the last
term avoids the double counting arising in the first term. The free
energy density is clearly a function of parameters
$\{m_{i\rightarrow\mu},\hat{m}_{\mu\rightarrow i}\}$, whose values
should make the free energy density stationary. In other words, the
recursive equations for $m_{i\rightarrow\mu},\hat{m}_{\mu\rightarrow
i}$ can be derived from a variational principle~\cite{cavity-2001}.
The result is summarized as follows:
\begin{subequations}\label{BP}
\begin{align}
m_{i\rightarrow\mu}&=\frac{e^{\beta h_{i}}\prod_{\nu\in\partial i\backslash\mu}\left[1+\hat{m}_{\nu\rightarrow i}\right]-e^{-\beta h_{i}}\prod_{\nu\in\partial i\backslash\mu}\left[1-\hat{m}_{\nu\rightarrow i}\right]}
{e^{\beta h_{i}}\prod_{\nu\in\partial i\backslash\mu}\left[1+\hat{m}_{\nu\rightarrow i}\right]+e^{-\beta h_{i}}\prod_{\nu\in\partial i\backslash\mu}\left[1-\hat{m}_{\nu\rightarrow i}\right]},\label{BPa}\\
\hat{m}_{\mu\rightarrow i}&=\prod_{j\in\partial\mu\backslash
i}m_{j\rightarrow\mu},
\end{align}
\end{subequations}
where $\partial\mu\backslash i$ denotes the neighbors of check $\mu$
with $i$ excluded, while $\partial i\backslash\mu$ denotes the
neighbors of bit $i$ with $\mu$ excluded. Once the iteration of
Eq.~(\ref{BP}) reaches a fixed point, the free energy contributions
$\Delta f_{i}$ and $\Delta f_{\mu}$ can be computed as:
\begin{subequations}\label{Contri}
\begin{align}
-\beta\Delta f_{i}&=\ln\left[e^{\beta h_{i}}\prod_{\mu\in\partial i}\frac{1+\hat{m}_{\mu\rightarrow i}}{2}+e^{-\beta h_{i}}\prod_{\mu\in\partial i}\frac{1-\hat{m}_{\mu\rightarrow i}}{2}\right],\\
-\beta\Delta
f_{\mu}&=\ln\left[\frac{1+\prod_{i\in\partial\mu}m_{i\rightarrow\mu}}{2}\right].
\end{align}
\end{subequations}
The magnetization $m_{i}$ in Eq.~(\ref{entro}) can be computed using
Eq.~(\ref{BPa}) with $\mu$ included. The above formulae can be
applied on single instances.

To study the typical property (e.g. computing typical entropy value)
of the problem, one should carry out the average over the quenched
disorder (channel statistics Eq.~(\ref{BSC}) and code
constructions). The free energy density is then given by
\begin{equation}\label{Pop}
    f=\sum_{l}\Lambda_{l}\left<\Delta f_{i}\right>_{{\rm pop}}-\frac{\left<l\right>}{\left<k\right>}\sum_{k}\mathcal
    {P}_{k}(k-1)\left<\Delta f_{\mu}\right>_{{\rm pop}},
\end{equation}
where the subscript ${\rm pop}$ means that the quantity is computed
from a population dynamics procedure. In practice, one starts from
an initial population of $\{m_{i\rightarrow\mu}\}$ of size $\mathcal
{N}$, whose elements are updated according to Eq.~(\ref{BP}) until a
stationary population is reached. The average is then carried out
using this stationary population which is detected if the free
energy density yields small fluctuation across iterations. Note that
the edge-perspective degree distribution should be used when the
incoming magnetizations $\{m_{j\rightarrow\mu}\}$ (or conjugated
magnetizations $\{\hat{m}_{\nu\rightarrow i}\}$) are sampled in
Eq.~(\ref{BP}).

The decoding performance can be evaluated by the decoding overlap:
\begin{equation}\label{olap}
    \rho=\frac{1}{M}\left<\sum_{i}\zeta_{i}{\rm
    sgn}\bigl<\sigma_{i}\bigr>_{\beta}\right>=\int dm\phi(m){\rm sgn}(m),
\end{equation}
where the gauge transformation has been performed and the inner
average is the thermal average, while the outer average is the
disorder one.
$\phi(m)=\sum_{l}\Lambda_{l}\int\prod_{\nu=1}^{l}P_{\nu}(\hat{m}_{\nu})\left<\delta(m-\mathcal
{F}(\{\hat{m}_{\nu}\}))\right>$ where $P_{\nu}$ represents the
distribution of incoming conjugated magnetization in the population
dynamics, and the average is taken with respect to the channel
statistics. $\mathcal {F}(\cdot)$ is given by the right hand side of
Eq.~(\ref{BPa}) including contributions from all neighbors of a bit
node.

\subsection{Zero temperature decoding: $1$RSB computation}
\label{subsec:ZTD}

The stability of iterations of Eq.~(\ref{BP}) depends on the
temperature and the noise level, which can be checked by adding a
small perturbation to the cavity magnetization on each edge, and
then updating the cavity magnetization and this additional
variance~\cite{Marinari-2006}, denoted by $v_{i\rightarrow\mu}$
whose evolution follows $v_{i\rightarrow\mu}=\sum_{\nu\in\partial
i\backslash\mu}\sum_{j\in\partial\nu\backslash
i}\left(\frac{\partial m_{i\rightarrow\mu}}{\partial
m_{j\rightarrow\nu}}\right)^{2}v_{j\rightarrow\nu}$. If the total
strength of the variances on all edges grows with the iteration, the
decorrelation assumption to derive Eq.~(\ref{BP}) breaks down,
implying that long-range correlation sets in among all nodes on the
graph. This does happen in the presence of low enough temperature
and high enough noise level. Therefore, we should consider one-step
replica symmetry breaking assumption when $T=0$, which concentrates
the Gibbs measure defined in Eq.~(\ref{poster}) on the ground state
configurations.

Under $1$RSB ansatz, the state space splits into an exponential
number of macroscopic states. Each state has its own free energy
density $f_{\alpha}$ with $\alpha$ being the state index. The Gibbs
measure correspondingly decomposes into contributions of various
free energy densities~\cite{Monasson-1995prl}, which is described by
introducing a generalized free energy function $\Phi(y)$:
\begin{equation}\label{Gfree}
    e^{M\Phi(y)}=\sum_{\alpha}e^{-yMf_{\alpha}}=\int d\epsilon
    e^{M(\Sigma(\epsilon)-y\epsilon)},
\end{equation}
where the complexity $\Sigma(\epsilon)$ is the growth rate function
of the number of states with the code length (system size)
$M$~\cite{cavity-2003}. The inverse-temperature-like parameter $y$
is used to fix the free energy density (energy density $\epsilon$ in
the zero temperature limit) of state, similar to the fact that the
temperature is used to select the energy of configurations. Using
definition Eq.~(\ref{Gfree}), one can derive the generalized free
energy contribution of a bit node $i$
($\Delta\Phi_{i}=\ln\left<e^{-y\Delta f_{i}}\right>$) and that of a
function node $\mu$ ($\Delta\Phi_{\mu}=\ln\left<e^{-y\Delta
f_{\mu}}\right>$), where the average is taken under $1$RSB
approximation. Note that in Eq.~(\ref{Gfree}), we have taken the
zero temperature limit while keeping a finite value of
$y$~\cite{Franz-2002pre}. We denote
$m_{i\rightarrow\mu}\equiv\tanh\beta h_{i\rightarrow\mu}$ and
$\hat{m}_{\mu\rightarrow i}\equiv\tanh\beta u_{\mu\rightarrow i}$.
The free energy contributions $\Delta f_{i}$ and $\Delta f_{\mu}$
are obtained from Eq.~(\ref{Contri}) in the limit
$\beta\rightarrow\infty$:
\begin{subequations}\label{Contri02}
\begin{align}
\Delta f_{i}&=-\left|h_{i}+\sum_{\mu\in\partial i}u_{\mu\rightarrow i}\right|+\sum_{\mu\in\partial i}|u_{\mu\rightarrow i}|,\\
\Delta
f_{\mu}&=2\Theta\left(-\prod_{j\in\partial\mu}h_{j\rightarrow\mu}\right)\min\left(|h_{j\rightarrow
\mu}|,j\in\partial\mu\right),
\end{align}
\end{subequations}
where $\Theta(x)$ is a step function taking values $\Theta(x)=0$ for
$x\leq 0$, $\Theta(x)=1$ for $x>0$. The distribution of
$\{h_{i\rightarrow\mu},u_{\mu\rightarrow i}\}$ satisfies the
following $1$RSB equation~\cite{cavity-2003}:
\begin{widetext}
\begin{subequations}\label{RSBeq}
\begin{align}
  P(h_{i\rightarrow\mu})\propto\int\left[\prod_{\nu\in\partial i\backslash\mu}du_{\nu\rightarrow i}
  Q(u_{\nu\rightarrow i})\right]e^{-y\Delta f_{i\rightarrow\mu}}\delta\left(h_{i\rightarrow\mu}-\mathcal {F}_{h}\bigl(\{u_{\nu\rightarrow i}\}\bigr)\right),\label{RSBeq01} \\
  Q(u_{\mu\rightarrow i})=\int\left[\prod_{j\in\partial\mu\backslash i}dh_{j\rightarrow\mu}
  P(h_{j\rightarrow\mu})\right]\delta\left(u_{\mu\rightarrow i}-\mathcal {F}_{u}\bigl(\{h_{j\rightarrow\mu}\}\bigr)\right),\label{RSBeq02}
  \end{align}
\end{subequations}
\end{widetext}
where $\mathcal {F}_{h}$ and $\mathcal {F}_{u}$ represent the zero
temperature limit of Eq.~(\ref{BP}) and the explicit form is given
in Sec.~\ref{subsec:Rbp}. The reweighting factor $e^{-y\Delta
f_{i\rightarrow\mu}}$ takes into account the reshuffling of free
energies of different states when an edge $i\rightarrow\mu$ is
removed~\cite{Battaglia-2004}. This factor intuitively discourages
(large) positive free energy change due to this cavity operation.
The $1$RSB equation~(\ref{RSBeq}) can be solved through a population
dynamics procedure~\cite{cavity-2001}. We give the details of the
algorithm in Appendix~\ref{sec:appendix}. During the iteration of
Eq.~(\ref{RSBeq}), one can also calculate the generalized free
energy and other thermodynamic quantities such as complexity, free
energy. These quantities of interest can be computed by the
following Legendre transformation~\cite{Franz-2002pre}:
\begin{subequations}\label{LT}
\begin{align}
  \Phi(y)&=\sum_{l}\Lambda_{l}\left<\Delta\Phi_{i}\right>-\frac{\bigl<l\bigr>}{\bigl<k\bigr>}\sum_{k}\mathcal {P}_{k}(k-1)\left<\Delta \Phi_{\mu}\right>,\label{LT01} \\
  \Sigma(\epsilon)&=\Phi(y)+y\epsilon,\label{LT02}\\
  \epsilon&=-\frac{\partial\Phi(y)}{\partial y}.\label{LT03}
  \end{align}
\end{subequations}
When the number of iteration is sufficiently large, the stationary
value of the above thermodynamic quantities can be obtained by the
bootstrap method~\cite{Hartmann-2009}.

We finally remark that Eq.~(\ref{RSBeq}) could not be further
simplified to an efficient survey propagation
algorithm~\cite{MM-2009}, mainly due to the fact that the support of
the cavity field distribution is not a finite discrete set, and
consequently a time-consuming sampling is required even in
application to single instances~\cite{SP-06}. Furthermore, the
finite value of $y$ should also be optimized.

\subsection{Reinforced belief propagation}
\label{subsec:Rbp}

In this section, we give the belief propagation equations in the
limit of $\beta\rightarrow\infty$. Taking $\beta\rightarrow\infty$
in Eq.~(\ref{BP}) leads to the following recursive equation (recall
that $m_{i\rightarrow\mu}\equiv\tanh\beta h_{i\rightarrow\mu}$ and
$\hat{m}_{\mu\rightarrow i}\equiv\tanh\beta u_{\mu\rightarrow i}$):
\begin{subequations}\label{BPzt}
\begin{align}
h_{i\rightarrow\mu}&=h_{i}+\sum_{\nu\in\partial i\backslash\mu}u_{\nu\rightarrow i},\label{BPzta}\\
u_{\mu\rightarrow i}&={\rm
sgn}\left(\prod_{j\in\partial\mu\backslash
i}h_{j\rightarrow\mu}\right)\min\left(|h_{j\rightarrow
\mu}|,j\in\partial\mu\backslash i\right),
\end{align}
\end{subequations}
where ${\rm sgn}(x)=x/|x|$ for $x\neq0$ and ${\rm sgn}(0)=0$. One
can apply the above iteration on single instances to infer the true
noise vector $\boldsymbol{\zeta}$, as $\sigma_{i}={\rm sgn}(H_{i})$
in which $H_{i}=h_{i}+\sum_{\mu\in\partial i}u_{\mu\rightarrow i}$
and $\{u_{\mu\rightarrow i}\}$ should be the stationary value.
However, in the glassy phase, the iteration fails to converge or
converges to a suboptimal solution ($\rho<1$). To circumvent this
problem, one can alternatively use the reinforcement
strategy~\cite{Braunstein-07ieee,Zecchina-2008entropy,Braunstein-11pre}.
In this situation, the external field is not constant but keeps
being updated as $h_{i}\rightarrow h_{i}+{\rm sgn}(H_{i})\delta$ at
each step with probability $1-t^{-r}$, where $t$ is the iteration
step and $r$ is the updating rate. At each step, one can also check
if the decoding overlap equals to unity (perfect recovery of the
original message). With a properly chosen value of $(\delta,r)$, the
iteration would converge to the true noise vector within certain
maximal number of iterations. The optimal values for $\delta$ and
$r$ can be obtained from several trials of decoding experiments.
Note that in each experiment, both parameters take values of small
magnitude. We will present the numerical simulation results at $T=0$
in Sec.~\ref{sec:RBP}.

\begin{figure}
\centering
    \includegraphics[bb=19 18 292 218,width=0.7\textwidth]{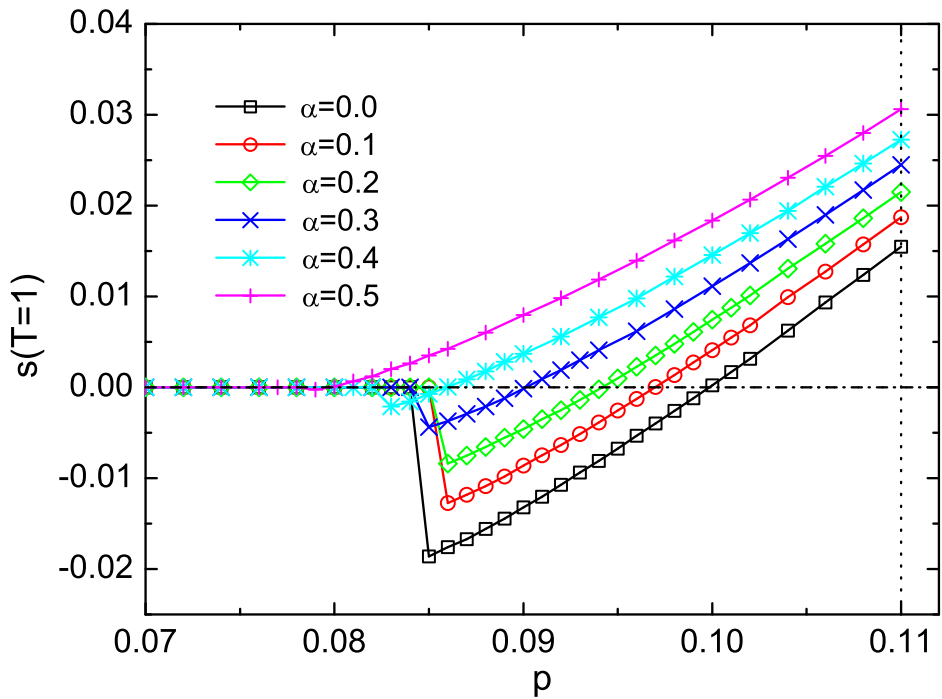}
  \caption{
  (Color online) Entropy density as a function of noise level $p$
  for different code constructions (code rate $R=0.5$). The temperature $T=1$. The
  vertical dotted line shows the Shannon's bound $p_{s}(R=0.5)\simeq0.110
  028$.
  }\label{Entropy}
\end{figure}

\section{Results and Discussions}
\label{sec_result}

In this section, we first present computation of entropy at the
Nishimori temperature $\beta=1$. $\beta=1$ corresponds to the
correct knowledge of the channel statistics characterized by the
noise level $p$. At this special temperature, the RS approximation
is correct and no further step replica symmetry breaking is
needed~\cite{Nishimori-1993,Montanari-2001}. Decoding at $T=1$ gives
the best performance among all temperatures~\cite{Nishimori-1993}.
In real situation, we do not know the true noise level of the noisy
channel. Then we can find alternatively the ground state
configuration during the decoding process, which entails the $1$RSB
analysis. The statistical mechanical analysis of zero temperature
decoding is presented in the third and fourth parts of this section.
Irregular and regular codes are both analyzed, in other words, we
assume $(\Lambda_{2},\Lambda_{3})=(\alpha,1-\alpha)$, $(\mathcal
{P}_{4},\mathcal {P}_{6})=(\alpha,1-\alpha)$ and other coefficients
vanish, such that the defined code ensemble has code rate $R=0.5$.
$\alpha=0$ corresponds to the regular code under consideration.

\subsection{Entropy at the Nishimori temperature}\label{sec_entro}

The entropy density at the Nishimori temperature is shown in
fig.~\ref{Entropy}. At the low noise level, there exists a
ferromagnetic state whose entropy vanishes and the decoding overlap
is always unity. The energy of the ferromagnetic state can be
computed as $f_{{\rm ferro}}=-(1-2p)F$ by setting $s=0$, and
$m_{i}=1$ $\forall i$~\cite{Montanari-2001}. As the noise level
increases up to a point where the entropy function starts to become
negative, metastable states with higher energy appear and compete
with the dominant ferromagnetic state. The normal belief propagation
would thus get stuck in one of these suboptimal metastable states.
In this sense, this transition point is called the dynamical
transition, denoted by $p_{d}$. The negativity of the entropy is due
to the emergence of metastable states, whose behavior as a function
of the energy density can be computed under $1$RSB ansatz and will
be shown later. Note that in this region, replica symmetry is still
correct for thermodynamically dominant state~\cite{Yoshida-2007}.
When the noise level reaches a point called static (thermodynamic)
transition point $p_{c}$, the entropy continuously becomes positive,
implying that the number of codewords degenerate with the true one
proliferates exponentially. Beyond the static transition, reliable
decoding is impossible, since multiple codewords dominate the state
space and one can not identify which one is the correct codeword.
$p_{c}$ evaluated at the Nishimori temperature coincides with that
of zero temperature decoding~\cite{Mourik-2002}, which was already
shown by assuming a frozen glassy phase~\cite{Montanari-2001}.

From fig.~\ref{Entropy}, one can also deduce that the irregular code
ensemble has a larger $p_{d}$, but increasing $\alpha$ makes the
separation between $p_{d}$ and $p_{c}$ smaller and smaller. It seems
that the irregular code with $\alpha=0.2$ has the largest $p_{d}$ of
all $\alpha$ shown in the figure. The superior performance of
irregular codes has also been found in similar
contexts~\cite{Vicente-2000,Richa-2001b,Luby-2001,Franz-2002pre},
but here we focus on the channel of BSC, which is different from the
case of  binary erasure channel for which analytical results can be
derived~\cite{Franz-2002pre}. A recent study found that for a chain
of multiple locally-coupled LDPC ensembles, $p_{d}$ of the coupled
system approaches $p_{c}$ of the original ensemble, which is called
threshold saturation via spatial coupling~\cite{Kudekar-2012}.

\begin{figure}
\centering
   \includegraphics[bb=14 14 294 216,width=0.7\textwidth]{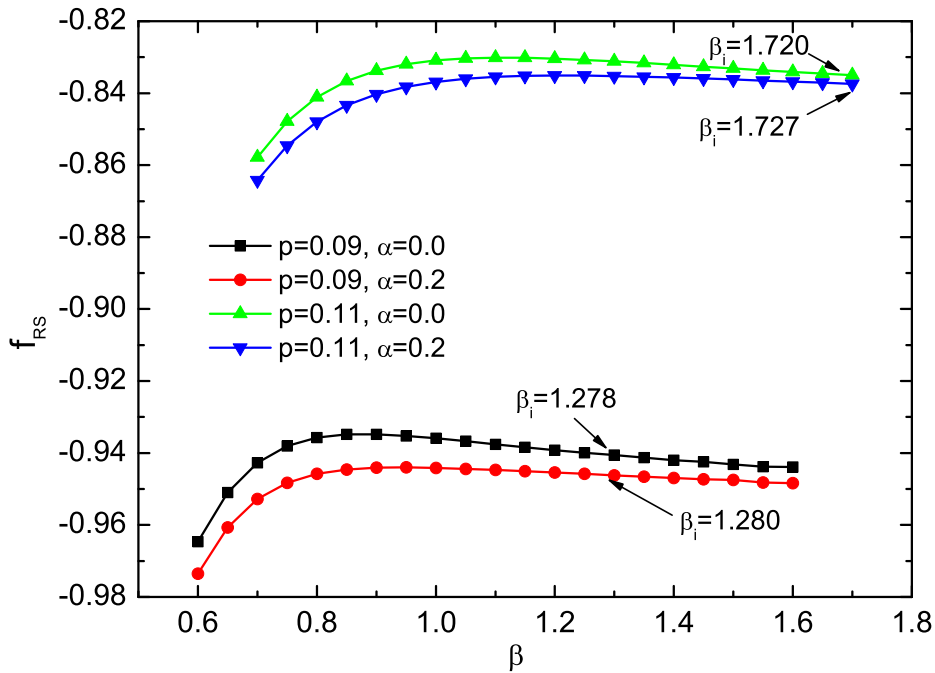}
  \caption{
  (Color online) Free energy density as a function of inverse temperature for regular ($\alpha=0.0$) and irregular ($\alpha=0.2$) codes.
 The curve develops a maximal value at a finite temperature $\beta_s^{-1}$
 where the entropy ($s=\beta^{2}\partial f_{{\rm
 RS}}/\partial\beta$) vanishes. Furthermore, the temperature where
 the RS ansatz becomes unstable is always lower than
 $\beta_{s}^{-1}$. For the curve from the top to the bottom, the
 instability temperature $\beta_{{\rm i}}$ is
 $1.720,1.727,1.278,1.280$ respectively.
  }\label{EntroCris}
\end{figure}
\subsection{The entropy crisis}
\label{sec:Ecri}

In fig.~\ref{EntroCris}, we show the free energy as a function of
inverse temperature. A maximum develops at a finite temperature
defined as $\beta_s^{-1}$. The stability analysis in
Sec.~\ref{subsec:ZTD} shows that $\beta_s$ is always smaller than
the instability inverse temperature $\beta_{{\rm i}}$, suggesting
that a discontinuous phase transition must appear at
$\beta_c\leq\beta_s$~\cite{Martin-2005} and a frozen glassy phase is
present. When calculating the complexity function, we find that the updating local
fields in Eq.~(\ref{RSBeq}) tend to diverge for larger values of
$y$, consistent with the frozen picture. The frozen glassy phase
implies that $1$RSB states are reduced to isolated configurations
with zero internal entropy, which is a direct result of the hard
constraints in Eq.~(\ref{Ham}) (the first term). Similar phenomenon
was also observed in the binary perceptron
problem~\cite{Huang-JPA2013} and other hard constraint satisfaction problems~\cite{Martin-2005}. For the decoding problem at the zero
temperature, the frozen picture is related to the difficulty for
local search heuristics to find the true sent message, since a
rearrangement of many bits is needed when changing one bit to
satisfy all parity-check constraints.

\subsection{Zero temperature decoding: improvement by reinforced belief propagation}
\label{sec:RBP}

\begin{figure}
\centering
    \includegraphics[bb=19 11 290 215,width=0.45\textwidth]{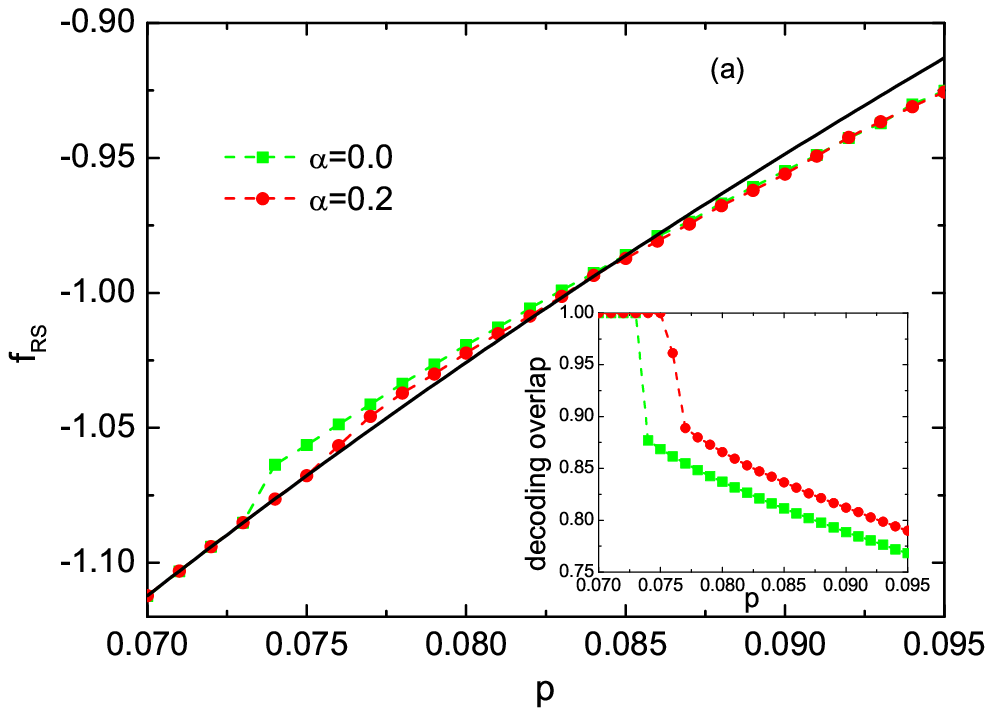}\hskip .1cm
    \includegraphics[bb=20 15 288 222,width=0.45\textwidth]{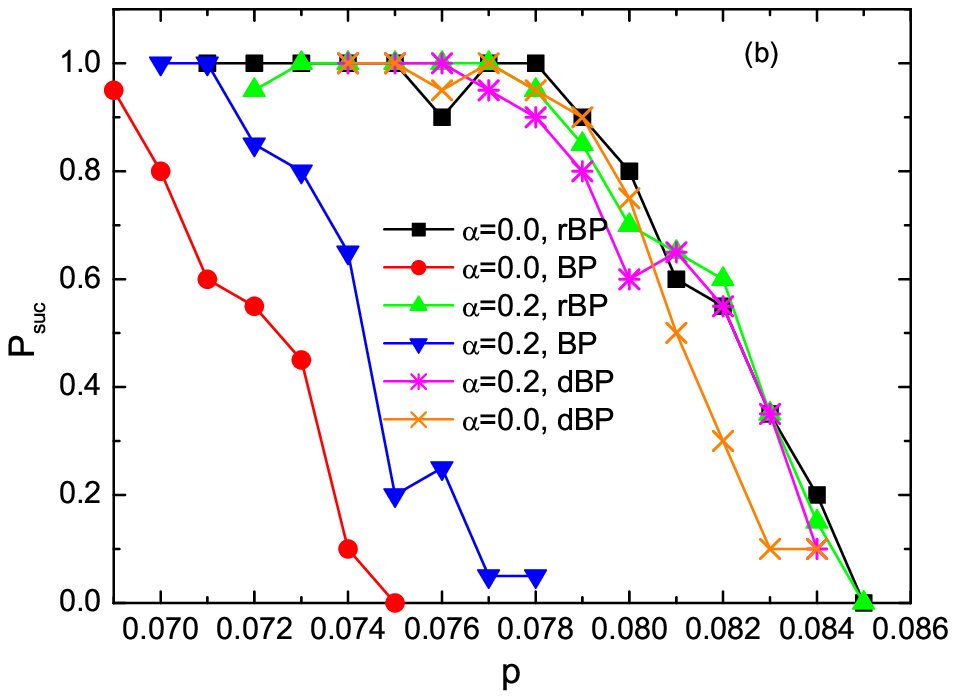}\vskip .1cm
    \includegraphics[bb=20 15 295 222,width=0.45\textwidth]{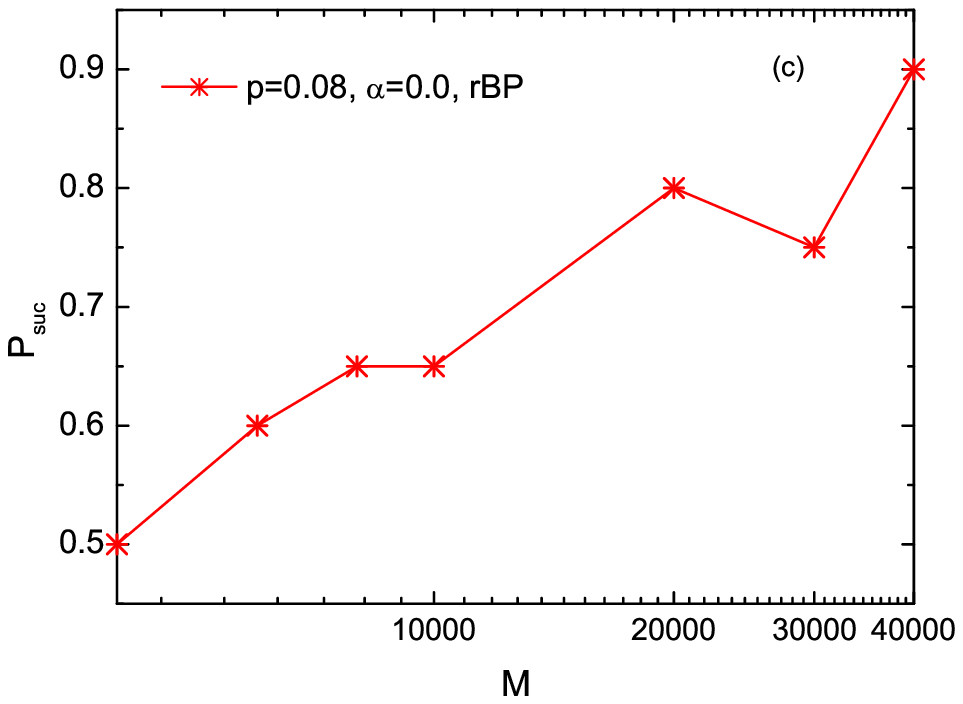}\vskip .1cm
  \caption{
  (Color online) (a) Phase transition for the zero temperature decoding.
  The full line represents the free energy of ferromagnetic state.
  The inset shows the transition in terms of decoding overlap. (b)
  Probability of successful decoding as a function of noise level.
  Performances of rBP, BP and dBP are compared. Code length $M=20000$.
  Code rate $R=0.5$ and the probability is computed over $20$ random
  samples. $(r,\delta)=(0.04,0.01)$ for rBP. The maximal number of iterations is equal to $1500$.
  (c) $P_{\rm suc}$ versus the code length $M$. Other parameters are
  the same as those in (b).
  }\label{ZTD}
\end{figure}

In this section, we first show the typical property of zero
temperature decoding, and then show an improvement of decoding
performance by applying reinforced belief propagation on single
instances. In fig.~\ref{ZTD} (a), we show the comparison between the
calculated free energy (Eqs.~(\ref{Pop}) and~(\ref{Contri02})) and
the ferromagnetic one. Before $p_{d}$, these two free energies
coincide, implying that the ferromagnetic state is the unique
dominant state without suboptimal metastable states. At $p_{d}$, the
calculated free energy jumps to a higher value, and keeps decreasing
as $p$ further increases, until the thermodynamic transition point
is reached. After the thermodynamic transition $p_c$, the calculated
free energy becomes lower than the ferromagnetic one, signalling
that a large number of codewords contribute to the Gibbs measure and
as a result, the ferromagnetic one ceases to be dominant. This
picture gives a rough estimate of the phase transition
points~\cite{Migliorini-2006}. More accurate determination requires
further steps of replica symmetry breaking, as analyzed in
Refs.~\cite{Franz-2002pre,Migliorini-2006}. In the inset of
fig.~\ref{ZTD} (a), we see clearly that the irregular code has
higher decoding threshold than the regular one, even when the
decoding is performed at the zero temperature.

Zero temperature decoding aims at finding the ground state
configuration as the inferred noise vector. As expected, normal
belief propagation (Eq.~(\ref{BPzt})) yields the same decoding
threshold as predicted in fig.~\ref{ZTD} (a). Surprisingly, the
decoding performances can be improved by applying an additional
updating external field to the normal BP iterations, the so-called
reinforcement strategy~\cite{Braunstein-07ieee,Braunstein-11pre}.
Due to emergence of metastable states with higher energy, normal BP
converges to these suboptimal states or fails to converge within a
maximal number of iterations. However, the reinforcement during the
iteration can bring the evolution of the passing messages (cavity
fields) on the links of factor graph to the desired codeword. The
decoding threshold can be pushed to a value as high as
$p\simeq0.082$ ($P_{{\rm suc}}=0.5$)~\cite{Vicente-2000}. This value
almost coincides with the theoretical decoding limit ($p_d$). It
should be mentioned that we also apply the same strategy to optimal
temperature decoding $(T=1)$, but it has little effects on the
decoding performances, i.e., yielding similar threshold with normal
BP. Thus, we conclude that, when the frozen glassy phase sets in,
both BP and rBP fail to recover the original message. However, by
using rBP, one can succeed in decoding at noise levels where normal
BP fails, as long as $p<p_d$, which is related to the fact that rBP
is more robust against complex energy landscape than BP in decoding
performance. The decoding performance of rBP versus the code length
is also shown in fig.~\ref{ZTD} (c). The finite size effect implies
that the successful decoding is achieved with high probability for
long-length codes as long as the noise level is below the threshold.

Finally, we remark that, by damping the updated message, i.e.,
$h_{i\rightarrow\mu}^{t+1}=\kappa
h_{i\rightarrow\mu}^{t+1}+(1-\kappa)h_{i\rightarrow\mu}^{t}$ (dBP),
where $\kappa$ is a small damping factor taking $0.05$ in our
simulations, one can also improve the decoding performance of BP
yielding comparable results with those of rBP (see fig.~\ref{ZTD}
(b)). However, the convergence of rBP to the true ferromagnetic
state is typically fast, e.g., rBP takes a median of $241$ iteration
steps, while dBP takes a median of $364$ iteration steps for
decoding the regular code of length $M=20000$ at $p=0.08$. Although
dBP performs slightly worse than rBP particularly around the
decoding threshold, they both improve greatly over the normal BP.
This is mainly due to the fact that, during the iteration, their
newly updated messages memorize the old messages at the preceding
step in different manners, suggesting that a memory of the history
of updating messages plays an important role in improving the
decoding performance especially when the complex energy landscape is
present.

\subsection{Zero temperature decoding: typical free-energy landscape under $1$RSB ansatz}
\label{sec:RSB}
\begin{figure}
\centering
    \includegraphics[bb=14 14 291 228,width=8.5cm]{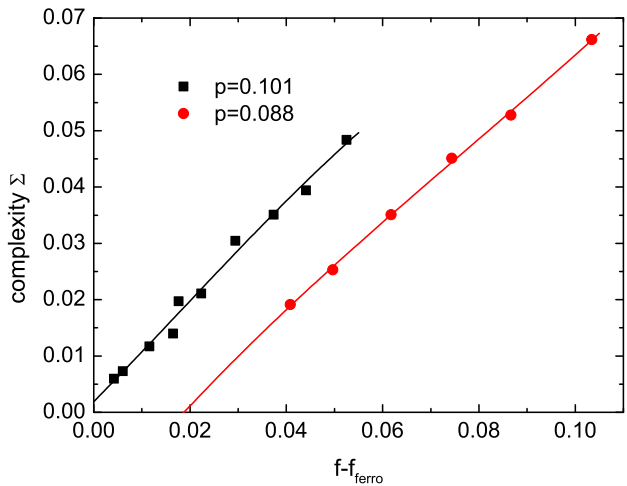}
  \caption{
  (Color online) Complexity as a function of free energy difference
  $f-f_{{\rm ferro}}$ for a regular code system ($\alpha=0$). Only the physical concave part of the curve
  is shown. $p=0.101\simeq p_{c}$ and $p=0.088>p_{d}\simeq0.084$.
  Curves are polynomial fits of degree $3$.
  }\label{complexity}
\end{figure}

In this section, we study the typical property of free energy
landscape of zero temperature decoding under $1$RSB ansatz, by
solving the $1$RSB equations derived in Sec.~\ref{subsec:ZTD}. Only
the regular code with $\alpha=0$ is considered, and qualitative
behavior is expected for irregular codes. We consider the complexity
as a function of the free energy difference $f-f_{{\rm ferro}}$. The
result is reported in fig.~\ref{complexity}. The population dynamics
details to solve the $1$RSB equations (Eq.~(\ref{RSBeq})) are
presented in Appendix~\ref{sec:appendix}. Varying the
inverse-temperature-like parameter $y$ from zero to positive value,
we first observe a convex unphysical part, followed by a concave
physical part which is shown in fig.~\ref{complexity} for $p=0.088$
and $p=0.101$. When $p$ is larger than $p_{d}$ but below $p_{c}$,
the complexity at certain $f>f_{{\rm ferro}}$ becomes positive,
demonstrating that the unique dominant state is the ferromagnetic
one, whereas, numerous high-lying metastable states are present, as
shown in the plot. These high-lying metastable states hide the true
ferromagnetic state, making the local search heuristics (such as
simulated annealing or normal BP) is difficult to find the desired
noise vector of the channel~\cite{Franz-2002pre}. $p_{d}$ is thus
defined as the point where a non-trivial concave part of the
complexity curve starts to appear. When the noise level becomes
larger than $p_{c}$, even at ferromagnetic free energy, the
complexity has a positive value, suggesting that the metastable
states have the same energy with the ferromagnetic state which is
not the unique one any more. Therefore decoding is impossible in
general due to the fact the the number of valid codewords grows
exponentially with the code length. $p_{c}$ is thus defined as the
point where the complexity at $f_{{\rm ferro}}$ starts to be
positive.

\section{Summary}
\label{sec_Con}

In this work, we provide a detailed statistical mechanics analysis
of low-density parity-check codes. The computation of entropy value
at the Nishimori temperature shows that one can improve the decoding
performance (shift the theoretical decoding threshold to a higher
value) by adopting the irregular code with optimal construction
parameters. The free energy function at different temperatures shows
that an entropy crisis occurs before the instability of RS
computation, which establishes the discontinuous nature of the phase
transition located at a temperature larger than or equal to the
entropy crisis temperature. Hard nature of the parity-check
constraints provides an intuitive understanding of the frozen
picture of codeword space structure. The divergence of cavity fields
observed in solving $1$RSB equations also supports this important
feature of LDPC codes.

At zero temperature, we find that the decoding performance can be
improved by applying a reinforced strategy during the iteration of
normal BP. Interestingly, this strategy has little effects on
temperature $T=1$ decoding which is expected to be optimal decoding
but the prior knowledge about the channel is required.  The normal
BP typically does not converge at high noise levels, indicating
$1$RSB approximation should be adopted to derive correct physical
picture. Under this ansatz, one can determine the dynamical
transition and thermodynamic transition by evaluating the complexity
function. Above the dynamical transition but below the thermodynamic
transition, the complexity becomes positive at certain higher level
of energy. For the higher noise level located above the static
transition point, the exponential growth of the number of degenerate
codeword is observed. Our studies provide a detailed quantitative
analysis of LDPC codes, in terms of code optimization, frozen glassy
phase and improved decoding algorithms when no prior knowledge of
channel statistics is assumed, under both RS and $1$RSB
approximations, and are expected to shed light on statistical
mechanics analysis of other state-of-the-art error correcting codes.

\section*{Acknowledgments}

This work was partially supported by the JSPS Fellowship for Foreign
Researchers (Grant No. $24\cdot02049$). Helpful discussions with Yoshiyuki Kabashima are acknowledged.
\appendix
\section{Population dynamics procedure solving $1$RSB equations}
\label{sec:appendix} In $1$RSB assumption, there exists a
distribution of cavity fields on each edge of a general factor
graph, capturing the fluctuation of cavity fields across different
macroscopic states. To take into account the graph ensembles
(average over different code constructions), we need a population of
cavity fields with $\mathcal {N}\times\mathcal {M}$ elements, i.e.,
$\mathcal{N}$ subpopulations (each of them has size $\mathcal {M}$).
The RS case corresponds to $\mathcal {M}=1$, and we use $\mathcal
{N}=20000$. A single update of one subpopulation proceeds in the
following five steps:
\begin{enumerate}
  \item  Sample a degree of a node $i$ from $\lambda_{l}$, then for
  each neighbor $\mu$, sample its degree from $\rho_{k}$, finally
  select at random and uniformly
  $k-1$ different subpopulations representing distributions on its
  adjacent incoming edges $j\rightarrow\mu$ except $i\rightarrow\mu$.
  A total number
  $(l-1)(k-1)$ of subpopulations are selected. Set an initial weight
  $w_{0}=e^{-y\Delta f_{i\rightarrow\nu}^{0}}$.
  \item  Compute one element $h_{i\rightarrow\nu}$ using these
  selected subpopulations according to Eq.~(\ref{RSBeq01}), and
  calculate the cavity free energy $\Delta f_{i\rightarrow\nu}$ at
  the same time. This newly computed element is accepted if the new
  weight $w=e^{-y\Delta f_{i\rightarrow\nu}}>w_{0}$, otherwise the
  old value is retained with a probability $1-w/w_0$.
  \item  Repeat (2) for each element of the subpopulation on edge
  $i\rightarrow\nu$ totally $\mathcal {L}$ times (we call $\mathcal
  {L}$ the sampling interval).
  \item  Sample a degree of a node $i$ from $\Lambda_{l}$, then for
  each neighbor $\mu$, sample its degree from $\rho_{k}$, finally
  select at random and uniformly
  $k-1$ different subpopulations representing distributions on its
  adjacent incoming edges $j\rightarrow\mu$ except $i\rightarrow\mu$.
  A total number
  $(k-1)l$ of subpopulations are selected. Using these
  subpopulations, one can compute $\left<\Delta\Phi_{i}\right>$ by
  repeating totally $\mathcal {L}\mathcal {M}$ sampling procedures.
   \item  Sample a degree of a function node $\mu$ from $\mathcal
   {P}_{k}$ and repeat step (1) to (4) $k$ times to get $k$ new
   subpopulations, which can be used to evaluate $\left<\Delta
   \Phi_{\mu}\right>$ using $\mathcal {L}\mathcal {M}$ samples.
\end{enumerate}

\begin{figure}
\centering
    \includegraphics[bb=18 13 288 220,width=8.5cm]{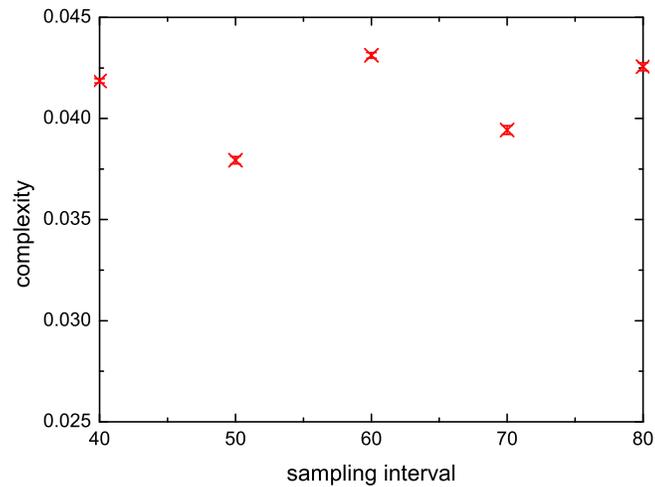}
  \caption{
  (Color online) Complexity as a function of sampling interval for a regular code system ($\alpha=0$).
   $y=0.82$ for $p=0.101$.
  }\label{SI}
\end{figure}

In practice, the above procedure is iterated for $\mathcal {T}$
steps (in unit of $\mathcal {N}$) with the first $\mathcal {T}_{0}$
steps discarded. The generalized free energy in Eq.~(\ref{LT01}) can
be obtained from the data of the later $\mathcal {T}-\mathcal
{T}_{0}$ iterations. In fact, the reweighting process is carried out
by using the Metropolis importance-sampling method~\cite{Zhou-2008pre}.
Other techniques can be found in the book~\cite{MM-2009}. The
parameters for the above population dynamics procedure are
$(\mathcal {N},\mathcal {M},\mathcal {L})=(1024,1024,70)$ and
$(\mathcal {T},\mathcal {T}_{0})=(400,200)$. Effects of $\mathcal
{L}$ on the complexity function are summarized in fig.~\ref{SI},
which shows that a sufficiently large $\mathcal{L}$ should be chosen
to ensure the reliable evaluation of relevant thermodynamic quantities.

\end{document}